\documentclass[11pt,a4 paper]{article}

\usepackage{amssymb}
\usepackage{amsmath}
\usepackage[dvips]{graphicx}
\begin{document}
\begin{center}
\noindent\textbf{\Large  VISCOUS FRW MODELS WITH PARTICLE CREATION IN EARLY UNIVERSE }\\
\end{center}

\vspace{0.5cm}
\begin{center}
\noindent  ${\bf C. P. Singh}$\\

\noindent {Department of Applied Mathematics,\\
 Delhi Technological university (Formerly Delhi College of
 Engineering),\\
 Bawana Road, Delhi-110 042, India.\\
 E-mail: cpsphd@rediffmail.com }\\
 \end{center}

\vspace{0.5cm}

\noindent \textbf{Abstract} We discuss the dynamical effects of bulk viscosity and particle creation on the early evolution of the Friedmann -Robertson -Walker model in the framework of open thermodynamical systems. We consider bulk viscosity and particle creation as separate irreversible processes. Exact solutions of the Einstein field equations are obtained by using the ``gamma-law" equation of state $p=(\gamma -1)\rho$, where the adiabatic parameter $\gamma$ varies with scale factor of the metric. We consider the cosmological model to study the early phases of the evolution of the universe as it goes from an inflationary phase to a radiation -dominated era in the presence of bulk viscosity and particle creation. Analytical solutions are obtained for particle number density and entropy for all models. It is found that, by choosing appropriate functions for particle creation rate and bulk viscous coefficient, the models exhibit singular and non-singular beginnings.\\


\noindent PACS number(s): 98.80.Cq; 98.80.-k; 04.20.Jb.\\

\pagebreak
\noindent\textbf{1. Introduction}\\

The study of particle creation in the relativistic cosmological models has drawn the attention of a number of authors. The first theoretical approach of the particle creation problem were investigated by Prigogine et al.$^{1, 2}$. They showed that the second law of thermodynamics may be modified to accommodate flow of energy from the gravitational field to the matter field, resulting in the creation of material particles. This leads to the reinterpretation of the stress tensor in general relativity, which now involves a time-asymmetric term depending on the rate of creation of particles. In other words, the process of particle creation out of gravitational energy is basically an irreversible phenomenon, capable of explaining the entropy burst in the expanding universe. A detailed study of the thermodynamics of the particle creation with changing specific entropy have been discussed by Lima et al.$^{3}$, Calv\~{a}o et al.$^{4}$, Lima and Germano $^{5}$, Lima et al.$^{6}$, and  Zimdahl et al.$^{7}$. Johri and Desikan $^{8}$ studied Friedmann-Robertson-Walker (FRW) models with particle creation. \\
\indent After the discovery of the accelerating universe the particle creation was reconsidered to explain it and got unexpected results. The particle creation pressure, which is negative, might play the role of dark energy component. Lima and Alcaniz $^{9}$, and Alcaniz and Lima $^{10}$ discussed FRW models with matter creation and analyzed the results through the observations. It was shown that the models with particle creation are consistent with the observational data. Zimdahl et al.$^{11}$ tested the particle creation with SNe Ia data and got the result of the accelerating universe. Yuan Qiang et al.$^{12}$ studied the models with adiabatic particle creation and showed that the model is consistent with SNe Ia data. \\
\indent On a phenomenological level particle creation has been described in the literature in terms of a bulk viscous stress. Padmanbhan and Chitre $^{13}$ discussed the role of the bulk viscosity in the entropy production in an expanding universe. Another peculiar characteristic of bulk viscosity is that it acts like a negative energy field in an expanding universe $^{14}$. The basic idea was that the bulk viscosity (particle creation) contributes at the level of the Einstein field equations as a negative pressure term. Barrow $^{15, 16}$ introduced this idea in the framework of new inflationary scenario. However, Prigogine et al.$^{1, 2}$ pointed out that the bulk viscosity and particle creation are not only two independent processes but, in general, lead to different histories of the evolution of the universe. Sudharsan and Johri $^{17}$, Brevik and Stokkan $^{18}$, and Triginer and pav\'{o}n $^{19}$ discussed the effect of bulk viscosity on the cosmological evolution of open thermodynamic systems which allow for simultaneous particle creation and entropy production. Their investigations reveal that the production of specific entropy is independent of the nature of the creation rate and depends only upon the coefficient of the bulk viscosity. This suggests that the bulk viscosity and particle creation are not only independent processes, but in general they lead to different histories of cosmic evolution. Desikan $^{20}$, Singh and Beesham $^{21}$, Johri and Pandey ${22}$, Singh et al.$^{23}$, and Singh and Kale $^{24}$ have studied the role of particle creation and bulk viscosity in isotropic and anisotropic models.\\
\indent It has been suggested in the literature that dissipative processes in the early stage of the expansion of the universe may well account for the present high degree of isotropy. A detail study of string -driven inflationary universe in terms of effective bulk viscosity coefficient was studied by Barrow $^{16}$. Gr{\o}n $^{25}$ and Maartens $^{26}$ presented exhaustive review on cosmological models with non-causal and causal thermodynamics respectively. Several authors $^{27-29}$ suggested that the bulk viscosity can drive the universe into a period of exponential expansion (inflation). This is really the case, as the effect of bulk viscosity in an expanding universe is to decrease the pressure making the total pressure negative. Chimento et al.$^{30}$ found that a mixture of a minimally coupled self-interacting scalar field and a perfect fluid is unable to drive the accelerated expansion, while a mixture of dissipative fluid with bulk viscosity and a minimally coupled self-interacting scalar field can successfully drive an accelerated expansion.\\
\indent In the standard model, the history of the universe begins with the radiation phase and then evolves to the present matter -dominated era. In order to overcome some of the difficulties met by standard model, Guth $^{31}$ proposed an inflationary phase and this would happen prior to the radiation -dominated phase. In general, the field equations are solved separately for the different epochs. However, Some authors have tried to solve the field equations in a unified manner. Madsen and Ellis $^{32}$ presented the evolution of the universe for inflationary, radiation and matter-dominated phases in a unified manner by assuming gamma $(\gamma)$ of ``gamma-law" equation of state $p=(\gamma -1)\rho$ as a function of scale factor of the FRW metric. Later on, Israelit and Rosen $^{33, 34}$ used a different equation of state to describe the transition from pre-matter to radiation and then radiation to matter-dominated phase in a unified manner. \\
\indent In a similar way, Carvalho $^{35}$ studied flat Friedmann-Robertson-Walker (FRW) model in general relativity by using the ``gamma-law" equation of state where $\gamma$ varies with cosmic time to describe the early phases (inflation and radiation) of the evolution of the universe in a unified manner. Therefore, it it not realistic to assume $\gamma$ as a constant throughout the history of the universe. We can obtain a reasonably realistic model if we assume the universe evolves through the epoches each of which $\gamma$ is constant. Singh et al.$^{36}$, Singh $^{37, 38}$ studied flat viscous FRW model and showed that the viscous fluid might drive the present acceleration of the universe. Recently, Singh and Beesham $^{39}$ have studied FRW model with the particle creation for the early phases of the evolution of the universe and discussed the kinematic tests. \\
\indent The aim of this paper is to extend Carvalho's work $^{35}$ to include the theory of particle creation and viscosity. We discuss the dynamical effects of bulk viscosity and particle creation on the early evolution of the Friedmann -Robertson -Walker model in the framework of open thermodynamical systems. We consider cosmological model with bulk viscosity and Particle creation as separate irreversible processes. Exact solutions of the Einstein field equations are obtained by using the ``gamma-law" equation of state $p=(\gamma -1)\rho$, where the adiabatic parameter $\gamma$ varies with scale factor of the metric. We consider the cosmological model to study the evolution of the universe as it goes from an inflationary phase to a radiation -dominated era in the presence of bulk viscosity and particle creation. Analytical solutions are obtained for particle number density and entropy for all models. It is found that, by choosing appropriate functions for particle creation and bulk viscous coefficient, the models exhibit singular and non-singular beginnings.\\

\noindent \textbf{2. Basic equations}\\

 We start with the homogeneous and isotropic flat Freidmann-Robertson-Walker (FRW) line element in the units $8\pi G=1$ and $c=1$
\begin{equation}
ds^{2}=dt^{2}-R^{2}(t)[dr^{2}+r^{2}(d\theta^{2}+sin^{2}\theta d\phi^{2})],
\end{equation}
\noindent where $r$, $\theta$, and $\phi$ are dimensionless comoving coordinates and $R$ is the scale factor.
The Einstein's field equations are given by
\begin{equation}
R_{ij}-\frac{1}{2}g_{ij}R= T_{ij},
\end{equation}
\noindent where $T_{ij}$ is the effective energy momentum tensor of the cosmic fluid in the presence of creation of particle and bulk viscosity, which includes the creation pressure term, $p_{c}$ and the bulk viscous stress, $\Pi$ and may be defined as $^{20}$
\begin{equation}
T_{ij}=(\rho+p+p_{c}+\Pi)u_{i}u_{j}-(p+p_{c}+\Pi)g_{ij}
\end{equation}
 \noindent where $\rho$ is the energy density, $p$ is thermodynamical pressure and $u_{i}$ is the four velocity vector satisfying the relation $u^{i}u_{i}=-1$. The creation pressure, $p_{c}$ is associated with the creation of particle out of the gravitational field $^{1, 2}$. The bulk viscous pressure, $\Pi$ represents only a small correction to the thermodynamical pressure, it is a reasonable assumption that the inclusion of viscous term in the energy -momentum tensor does not change fundamentally the dynamics of the cosmic evolution. \\
\indent In context of `open' system with adiabatic creation, the non-trivial Einstein's field equations for a fluid endowed with matter creation  and viscosity can be written as
\begin{equation}
3\frac{\dot{R^{2}}}{{R}^{2}}= \rho,
\end{equation}
\begin{equation}
2\frac{\ddot{R}}{R}+\frac{\dot{R^{2}}}{{R}^{2}}=-(p+p_{c}+\Pi),
\end{equation}
\noindent where an overhead dot denotes derivative with respect to cosmic time $t$. For adiabatic particle creation, the pressure $p_{c}$ assumes the following form $^{4, 5, 20}$
\begin{equation}
p_{c}=-\frac{(\rho+p)V}{N}\frac{dN}{dV},
\end{equation}
\noindent where $V=R^{3}$ is the 3-space volume and $N$ is the particle number. In models with adiabatic creation, the balance equation for the particle number density $n$, where $n= N/V$ is given by $^{4, 20}$
\begin{equation}
\frac{\dot{N}}{N}=\frac{\dot{n}}{n}+3\frac{\dot{R}}{R}=\frac{\Gamma(t)}{n},
\end{equation}
\noindent where $\Gamma(t)$ denotes a source term, i.e.,  the particle creation rate which will be positive ($\Gamma > 0$) or negative ($\Gamma < 0 $) depending on whether there is production or annihilation of particles. \\
\indent In this work we take the simple phenomenological expression of particle creation rate $^{6}$
\begin{equation}
\Gamma(t)=3\beta n H,
\end{equation}
\noindent where the parameter $\beta$ is defined on the interval [0, 1], which is assumed to be constant and $H=\dot{R}/{R}=(1/3)(\dot{V}/V)$ is the Hubble parameter. \\
Using (7) and (8) into (6), the particle creation pressure reduces to
\begin{equation}
p_{c}=-\beta(\rho+p).
\end{equation}
Equations (4) and (5) lead to the continuity equation
\begin{equation}
\dot{\rho}+3(\rho+p)H=-3(p_{c}+\Pi)H.
\end{equation}
\indent The conventional bulk viscous effect in a FRW universe can be modelled within the framework of non-equilibrium thermodynamics proposed by Israel and Stewart$^{40}$. In this theory, the transport equation for the bulk viscous pressure $\Pi$ takes the form
\begin{equation}
\Pi+\tau \dot{\Pi}=-3\zeta H-\frac{\tau \Pi}{2}\left[3H+\frac{\dot{\Pi}}{\Pi}-\frac{\dot{T}}{T}-\frac{\dot{\zeta}}{\zeta}\right],
\end{equation}
\noindent where the positive definite quantity $\zeta$ stands for the coefficient of the bulk viscosity, $T$ is the temperature of the fluid, and $\tau$ is the relaxation time associated with the dissipative effect.
\noindent We consider here the role of bulk viscosity in framework of equilibrium thermodynamics. Although equilibrium thermodynamics has some shortcomings, in contrast to the extended irreversible thermodynamics, the treatment of bulk viscosity under equilibrium thermodynamics in the present paper is justified for the following reasons.\\
\indent Provided the factor in the square bracket is small, one can approximate the equation (11) as a simple form
\begin{equation}
\Pi+\tau \dot{\Pi}=-3\zeta H,
\end{equation}
Assuming $\tau=\zeta/\rho$ as taken by Maartens $^{26}$, and noting that the energy density $\rho$ is proportional to the square of the scalar expansion $\theta$, we have $\Pi=-\zeta \theta [1+\dot{\Pi}/\theta^{3}]$. It is seen that we can take the equilibrium value for $\Pi$, namely,
\begin{equation}
\Pi=-3\zeta H,
\end{equation}
\noindent provided $\dot{\Pi}/\theta^{3}\ll 1$. In the particle creation models, $\theta$ happens to be a rapidly increasing function of time, and hence the condition  $\dot{\Pi}/\theta^{3}\ll 1$ holds.  The entropy equation is assumed to be of the form
\begin{equation}
T \frac{\dot{S}}{V}=\zeta \theta^{2}+T\frac{S}{V}\frac{\dot{N}}{N},
\end{equation}
\noindent where $S$ is the entropy and $\theta=3(\dot{R}/R)$ is the expansion scalar. In terms of specific entropy per particle, $\sigma$, Eq.(14) can be written as
\begin{equation}
\dot{\sigma}=\frac{\zeta \theta^{2}}{Tn}
\end{equation}
We see that the production of specific entropy per particle is independent of the nature of $\Gamma(t)$ and depends only on the nature of bulk viscosity. If $\zeta=0$, we have $\sigma$ =constant.\\

\noindent \textbf{3. Solution of field equations}\\

In order to solve the field equations, we suppose that the pressure $p$ and energy density $\rho$ are related through the ``gamma-law" equation of state
\begin{equation}
p=(\gamma-1)\rho.
\end{equation}
\indent In general, the value of $\gamma$ is taken to be constant and lying in the interval $0\leq \gamma \leq 2$. But our aim in this paper is to let the parameter $\gamma$ depends on scale factor $R$ to describe the early phases, inflationary and radiation -dominated  evolution of the universe in a unified manner. We assume, following  Carvalho $^{35}$ that, the functional form of $\gamma$ as
\begin{equation}
\gamma(R)=\frac{4}{3}\frac{A(R/R_{*})^{2}+(a/2)(R/R_{*})^{a}}{A(R/R_{*})^{2}+(R/R_{*})^{a}},
\end{equation}
\noindent where $A$ is a constant and $`a'$ is free parameter related to the power of the cosmic time $t$ during the inflationary phase. Here, $R_{*}$ is a certain reference value of $R$. The function $\gamma(R)$ is defined in such a manner that when the scale factor $R$ is less than $R_{*}$, i.e., when $R < < R_{*}$, an inflationary phase ($\gamma \leq 2a/3$) can be obtained and for $R > > R_{*}$ we have a radiation-dominated phase ($\gamma=4/3$). The expression of $\gamma(R)$ in Eq.(17) is an increasing function of $R$. In the limit $R\rightarrow 0$, $\gamma(R)=\frac{2a}{3}$. Thus, $1$ is the maximum value of $`a'$ for an inflation epoch to exist. As $`a' $ approaches to zero we have an exponential inflation ($\gamma=0$). Therefore, $a$ must lie in the interval $0\leq a < 1$.\\
\indent Using (9), (13) and (16) into (10), we find
\begin{equation}
\dot{\rho}+3\gamma (1-\beta)\rho H=9\zeta H^{2},
\end{equation}
\noindent which can be written as
\begin{equation}
\frac{\rho^{\prime}}{\rho}+3(1-\beta)\frac{\gamma(R)}{R}=\frac{9\zeta H}{\rho R},
\end{equation}
\noindent where a prime denotes derivative with respect to the scale factor $R$.\\
Equation (4) can be rewritten as
\begin{equation}
\frac{\rho^{\prime}}{\rho}=\frac{2H^{\prime}}{H}.
\end{equation}
Using (4) and (20) into (19), we finally get
\begin{equation}
H^{\prime}+\frac{3}{2}(1-\beta)\gamma(R)\frac{H}{R}=\frac{3}{2}\frac{\zeta}{R}.
\end{equation}
\noindent Equation (21), involving  $H$ and  $\zeta $, admits solution for  $H$  only if  $\zeta $ is specified. In the homogeneous models, $\zeta $ depends only on time and therefore we may consider it as a function of the universe energy density. According to literature developments [see refs., Belinskii and Khalatnikov $^{41}$, Barrow $^{15}$ and Maartens $^{26}$], we assume that the bulk viscosity coefficient depends on  $\rho $ via a power --law of the form \\
\begin{equation}
\zeta =\zeta _{0} \rho ^{m},
\end{equation}
\noindent where $\zeta _{0}$ is a non -negative constant and  $m(\ge 0)$ is numerical constant to be specified later. It is standard to assume the above law in the absence of better alternatives. In what follow we  solve Eq.(21) by taking the various physical assumptions on $m$\\

\noindent \textbf{3.1 Solution with $m=0$}\\

\noindent In this case, we get the constant coefficient of bulk viscosity, i.e., $\zeta=\zeta_{0}$ and therefore, Eq. (21) now reduces to  \\
\begin{equation}
H^{\prime}+\frac{3}{2}(1-\beta)\gamma(R)\frac{H}{R}=\frac{3}{2}\frac{\zeta_{0}}{R}.
\end{equation}
\noindent Substituting (17) into (23) and solving, we get
\begin{equation}
H \left[A(R/R_{*})^{2}+(R/R_{*})^{a}\right]^{(1-\beta)}=C_{0}+\frac{3\zeta_{0}}{2}\int \frac{[A(R/R_{*})^{2}+(R/R_{*})^{a}]^{(1-\beta)}}{R}dR
\end{equation}
\noindent where $C_{0}$ is a constant of integration. We now solve Eq.(24) for two early phases of the universe viz. inflationary and radiation -dominated phases, respectively.\\
\indent For inflationary phase $(R < < R_{*})$, Eq(24) gives
\begin{equation}
H=\frac{C_{0}}{(R/R_{*})^{a(1-\beta)}}+\frac{3\zeta_{0}}{2a(1-\beta)},
\end{equation}
\noindent and for radiation-dominated phase $(R > > R_{*})$, we have
\begin{equation}
H=\frac{C_{0}}{A^{(1-\beta)}(R/R_{*})^{2(1-\beta)}}+\frac{3\zeta_{0}}{4(1-\beta)},
\end{equation}
\noindent From above we find that the Hubble parameter is constant when the viscous  term dominates in both the phases and therefore the solutions are de -Sitter type where the scale factor varies as $R\propto \exp(3\zeta_{0}/2a(1-\beta))$ or  $R\propto \exp(3\zeta_{0}/4(1-\beta))$ and density is finite. The universe has no singularity. When the particle creation dominates we get the singular model as $R\propto t^{1/a(1-\beta)}$ or $R\propto t^{1/2(1-\beta)}$. \\
\indent When $C_{0}\neq 0$, after some mathematical manipulations between the constants , Eq. (25) gives
\begin{equation}
\left(\frac{R}{R_{*}}\right)^{a(1-\beta)}=\left[2a B\left\{\frac{\exp(\frac{3\zeta_{0}}{2}t)-(1-\beta)}{3\zeta_{0}}\right\}\right],
\end{equation}
\noindent where $B$ is another constant. The Hubble parameter has the form
\begin{equation}
H=\frac{3\zeta_{0}}{2a(1-\beta)}\left[1-(1-\beta)\exp\left(-\frac{3\zeta_{0}}{2}t\right)\right]^{-1}
\end{equation}
The energy density and particle pressure are respectively given by
\begin{equation}
\rho=\frac{27\zeta^{2}_{0}}{4a^{2}(1-\beta)^{2}}\left[1-(1-\beta)\exp\left(-\frac{3\zeta_{0}}{2}t\right)\right]^{-2},
\end{equation}
\begin{equation}
p_{c}=-\frac{9\beta\zeta^{2}_{0}}{2a(1-\beta)^{2}}\left[1-(1-\beta)\exp\left(-\frac{3\zeta_{0}}{2}t\right)\right]^{-2}.
\end{equation}
The particle number density is given by
\begin{equation}
n=n_{0i}\left(\frac{R}{R_{*}}\right)^{-3(1-\beta)}.
\end{equation}
\noindent We note that the effect of particle creation is measured by the parameter $\beta$. The particle number is given by
\begin{equation}
N=N_{0i}\left(\frac{R}{R_{*}}\right)^{3\beta}.
\end{equation}
\noindent In the above expressions the subscript ``$0i$" refers to the present observed values of the parameters during inflationary phase. It is observed that the universe starts from a non-singular state. We observe that for $\rho >0$, $\beta$ must be in the interval $0\leq \beta <1$, which is also the condition for the expansion of the universe. It shows that$\rho$ starts with a finite value at $t=0$ and ends up with a finite value at $t\rightarrow \infty$. It is interesting to note that the inflationary solution can be obtained with or without particle creation and this is due to constant bulk viscous coefficient. In the absence of both bulk viscosity and particle creation one get  the power-law inflation, i.e., $R\propto t^{1/a}$, which exhibits singular model. We also find that $N$ increases with $R$ increases. For $\beta=0$, $N$ would remain constant throughout the evolution of the universe.\\
\indent In order to get an expression for $\sigma$ we need an additional equation of state involving $T$. For the sake of simplicity we take the ideal gas equation $p=nk_{b}T$ with $k_{b}$, the Boltzmann constant, which we use just as an approximation since ideal gas lack bulk viscosity. Inserting this value into Eq.(15) for $\gamma=2a/3$ we get
\begin{equation}
\sigma=\frac{9\zeta_{0}k_{b}}{(2a-3)}t+l,
\end{equation}
\noindent where $l$ is an integration constant.\\
\indent On the other hand, when the universe is dominated by radiation, i.e., when $R > > R_{*}$, Eq.(26) gives
\begin{equation}
\left(\frac{R}{R_{*}}\right)^{2(1-\beta)}=\left[\frac{4 B_{1}}{A^{(1-\beta)}}\left\{\frac{\exp(\frac{3\zeta_{0}}{2}t)-(1-\beta)}{3\zeta_{0}}\right\}\right],
\end{equation}
\noindent where $B_{1}$ is a constant of integration. The Hubble parameter has the form
\begin{equation}
H=\frac{3\zeta_{0}}{4(1-\beta)}\left[1-(1-\beta)\exp\left(-\frac{3\zeta_{0}}{2}t\right)\right]^{-1}.
\end{equation}
The energy density and particle pressure are respectively given by
\begin{equation}
\rho=\frac{27\zeta^{2}_{0}}{16(1-\beta)^{2}}\left[1-(1-\beta)\exp\left(-\frac{3\zeta_{0}}{2}t\right)\right]^{-2},
\end{equation}
\begin{equation}
p_{c}=-\frac{9\beta\zeta^{2}_{0}}{4(1-\beta)^{2}}\left[1-(1-\beta)\exp\left(-\frac{3\zeta_{0}}{2}t\right)\right]^{-2}.
\end{equation}
The particle number density and correspondingly particle number are given by Eq.(31) and (32) where the scale factor is defined by Eq.(34). The specific entropy per particle is found to be
\begin{equation}
\sigma=9\zeta_{0}k_{b}t+l_{1},
\end{equation}
\noindent where $l_{1}$ is constant of integration.\\
\indent In this phase the physical interpretation is similar to the case of the inflationary phase. The bulk viscosity avoids the singularity. For the expansion of the universe, we must have $0\leq \beta < 1$. \\
\indent The deceleration parameter, which is defined as $q=-(R\ddot{R}/R^{2})$, varies from $q=[a(1-\beta)\exp(-3\zeta_{0}t/2)-1]$ for inflationary phase to $q=[2(1-\beta)\exp(-3\zeta_{0}t/2)-1]$ in radiation -dominated phase. We find that $q=[a(1-\beta)-1]$ in inflationary phase and  $q=(1-2\beta)$ in radiation phase as $t=0$. However, it is $q=-1$ in both phases as $t\rightarrow\infty$, which shows the inflation during the late times of evolution of the universe.\\
\indent Now, we study the model in the limit $a\rightarrow 0$ and in this case Eq. (24) becomes
\begin{equation}
H \left[A(R/R_{*})^{2}+1\right]^{(1-\beta)}=C_{0}+\frac{3\zeta_{0}}{2}\int \frac{[A(R/R_{*})^{2}+1]^{(1-\beta)}}{R}dR.
\end{equation}
\noindent Again, in the limit of very small $R$ $(R < < R_{*})$, we get
\begin{equation}
R=R_{*}\exp \left[\frac{\exp\left(\frac{3\zeta_{0}}{2}t\right)-C_{0}}{3\zeta_{0}/2}\right],
\end{equation}
\noindent which means that the universe has `superinflationary' expansion due to the presence of viscous term. The effect of particle creation is negligible. It has finite dimension as $t\rightarrow -\infty$. The Hubble parameter has the value $H=\exp(3\zeta_{0}t/2)$ which shows that the expansion is driven by viscosity. \\
\noindent The energy density is given by
\begin{equation}
\rho=3\exp(3\zeta_{0}t)
\end{equation}
\noindent The energy density is finite at $t=0$ and hence there is no physical singularity. These solutions help to solve several cosmological problems like flatness, horizon, monopole, etc. associated with standard model. The particle creation pressure is zero. The superinflation is due to either by the bulk viscous coefficient or vacuum energy density ($p=-\rho$). Thus the cosmological constant may also be considered to give rise the exponential inflation in the absence of creation pressure. We also observe that the particle density decreases whereas the number of particle increases exponentially during this phase.\\
\indent In the limit of very large R, when the universe enters to radiation-dominated phase, we obtain the same physical expressions for different physical parameters as in the case of $0 < a < 1$. We see that the deceleration parameter varies from $q=-1$ at $R=0$ to $q=(1-2\beta)$ for radiation-dominated phase.\\

\noindent \textbf{3.2 Solution with $m\neq 0$}\\

\noindent Using (4) and (22), Eq.(21) now reduces to
\begin{equation}
H^{\prime}+\frac{3}{2}(1-\beta)\gamma(R)\frac{H}{R}=\frac{3^{(m+1)}\zeta_{0}}{2}\frac{H^{2m}}{R}.
\end{equation}
\noindent Substituting (17) into (42) and integrating, we have the solution, for all  $m\ne 1/2$, which is given by  \\
\begin{eqnarray}
 \frac{1}{H^{(2m-1)} [A(R/R_{*} )^{2} +(R/R_{*} )^{a} ]^{(1-\beta)(2m-1)} }  \nonumber \\
= C_{1} -\frac{(2m-1)3^{m+1} \zeta _{0} }{2} \int \frac{dR}{R[A(R/R_{*} )^{2} +(R/R_{*} )^{a} ]^{(1-\beta)(2m-1)} },
\end{eqnarray}
\noindent where $C_{1}$ is a constant of integration and $0 < a < 1$. We discuss the solution for two early phases of the universe viz. inflationary and radiation-dominated phases respectively. \\
\noindent For inflationary phase ($R < < R_{*}$), the expression for Hubble parameter is given by
\begin{equation}
H^{(2m-1)}=\frac{1}{[C_{1}(R/R_{*})^{a(2m-1)(1-\beta)}+(3^{(m+1)}\zeta_{0}/2a(1-\beta))]}.
\end{equation}
The energy density, creation pressure and bulk viscous coefficient in terms of scale factor for this phase are respectively given by
\begin{equation}
\rho =3[C_{1}(R/R_{*} )^{a(2m-1)(1-\beta)} +(3^{m+1} \zeta _{0} /2a(1-\beta))]^{2/(1-2m)},
\end{equation}
\begin{equation}
p_{c} =-2a\beta[C_{1}(R/R_{*} )^{a(2m-1)(1-\beta)} +(3^{m+1} \zeta _{0} /2a(1-\beta))]^{2/(1-2m)},
\end{equation}
\begin{equation}
\zeta =3^{m}\zeta _{0} [C_{1}(R/R_{*})^{a(2m-1)(1-\beta)} +(3^{m+1} \zeta _{0} /2a(1-\beta))]^{2m/(1-2m)}.
\end{equation}
An important observational quantity is the deceleration parameter $q$. In the case of inflationary phase, the deceleration parameter can be expressed as a function of the scale factor as
\begin{equation}
q=\frac{(a(1-\beta)-1)C_{1}(R/R_{*} )^{a(2m-1)(1-\beta)} -(3^{m+1} \zeta _{0} /2a(1-\beta))}{[C_{1}(R/R_{*} )^{a(2m-1)(1-\beta)} +(3^{m+1} \zeta _{0} /2a(1-\beta))]}
\end{equation}
The sign of $q$ indicates whether the model inflates or not. The negative value of $q$ ($q < 0$) describes the acceleration of the universe whereas the positive one $(q > 0)$ describes the decelerating universe. In the absence of viscous term, it becomes $q= a(1-\beta)-1$. This shows that the universe decelerates for $a(1-\beta)> 1$ and accelerates for $a(1-\beta)<1$. Since $0 \leq a < 1$, in the absence of creation pressure, i.e., $\beta=0$ and $C_{1}\geq 0$, we get the negative value of $q$ showing the inflation in the model. \\
\noindent For radiation -dominated phase ($R > > R_{*}$), we have
\begin{equation}
H^{(2m-1)}=\frac{1}{[C_{1}A^{(2m-1)(1-\beta)}(R/R_{*})^{2(2m-1)(1-\beta)}+(3^{(m+1)}\zeta_{0}/4(1-\beta))]}.
\end{equation}
In this phase the energy density, particle pressure and bulk viscous coefficient in terms of scale factor are given by
\begin{equation}
\rho =3[C_{1}(R/R_{*} )^{2(2m-1)(1-\beta)} +(3^{m+1} \zeta _{0} /4(1-\beta))]^{2/(1-2m)},
\end{equation}
\begin{equation}
p_{c} =-4\beta[C_{1}(R/R_{*} )^{2(2m-1)(1-\beta)} +(3^{m+1} \zeta _{0} /4(1-\beta))]^{2/(1-2m)},
\end{equation}
\begin{equation}
\zeta =3^{m}\zeta _{0} [C_{1}(R/R_{*})^{2(2m-1)(1-\beta)} +(3^{m+1} \zeta _{0} /4(1-\beta))]^{2m/(1-2m)}.
\end{equation}
The deceleration parameter, in this case, in terms of scale factor can be written as
 \begin{equation}
 q=\frac{A^{(2m-1)(1-\beta)} (1-2\beta)C_{1}(R/R_{*} )^{2(2m-1)(1-\beta)} -(3^{m+1} \zeta _{0} /4(1-\beta))}{[A^{(2m-1)(1-\beta)} C_{1}(R/R_{*} )^{2(2m-1)(1-\beta)} +(3^{m+1} \zeta _{0} /4(1-\beta))]}
 \end{equation}
In the absence of viscous term, we have $q=(1-2\beta)$. This shows that the model decelerates for $0\leq \beta < 1/2$ and accelerates for $1/2 < \beta <1$ in the absence of viscous term. If $C_{1}> 0$, we observe that $q$ is positive for $(R/R_{*} )^{2(2m-1)(1-\beta)} >(3^{m+1} \zeta _{0} /4(1-\beta)A^{(2m-1)(1-\beta)} C_{1})$, $q$ is negative for $(R/R_{*} )^{2(2m-1)(1-\beta)} <(3^{m+1} \zeta _{0} /4(1-\beta)A^{(2m-1)(1-\beta)} C_{1})$ and $q=0$ for $(R/R_{*} )^{2(2m-1)(1-\beta)} =(3^{m+1} \zeta _{0} /4(1-\beta)A^{(2m-1)(1-\beta} C_{1})$. Again, if $C_{1}=0$, we get $q=-1$, which shows the acceleration of the universe.\\

\indent We discuss the more general solution of the above models for the following  three cases depending on the value of $m$ for inflationary and radiation -dominated phases.\\

\noindent \textbf{3.2.1 Solution with $ m > 1/2$}\\

In this case, we observe that Eqs. (44) and (49) give constant value of Hubble parameter as $R\rightarrow 0$. The model is similar to those of Murphy $^{27}$, which has initially the viscosity -dominated steady state behavior. The bulk viscosity appears to be the effective mechanism to remove the initial singularity. Let us discuss this case by taking $m=1$ as the bulk viscosity is proportional to the energy density. In this case, Eqs. (44) and (49) for inflationary and radiation -dominated phases, respectively can be explicitly obtained as
\begin{equation}
\frac{C_{1}}{a(1-\beta)} \left(\frac{R}{R_{*} } \right)^{a(1-\beta)} +\frac{9\zeta _{0} }{2a(1-\beta)} \ln R=t+t_{0}.
\end{equation}
\begin{equation}
\frac{C_{1}}{2(1-\beta)} \left(\frac{R}{R_{*} } \right)^{2(1-\beta)} +\frac{9\zeta _{0} }{4(1-\beta)} \ln R=t+t_{0}.
\end{equation}
\noindent where $t_{0} $ is the integration constant and can be adjusted to zero. In general, it is not possible to express  $R$  explicitly as a function of time. For sufficiently small $R$, the second term in equations (54) and (55) dominate over the first, containing the viscous term. Thus, we get exponential expansion of the form  $R=\exp (H_{*} t)$, where  $H_{*} =2a(1-\beta)/9\zeta _{0} $ or  $H_{*} =4(1-\beta)/9\zeta _{0} $, which represents the singularity free model and all solutions approach to the de Sitter state with the expansion rate  $H=H_{*} $. Such a non --singular behavior is exhibited only in the presence of bulk viscosity. Such type of solution has been obtained by Murphy (1973), where he attributed the viscosity effect to gravitons production in the graviton --gravitons scattering. Since $H_{*}\propto \zeta_{0}^{-1}$, so the smaller is the value of $\zeta_{0}$ the bigger is the inflation rate. During inflation both $\rho$ and $\zeta$ are finite. The deceleration parameter has the value  $q=-1$. The above solutions (54) and (55) can also be directly obtained  from the general solution (44) or (49) after some manipulation. \\
\indent On the other hand, if the first term dominates we get the power --law expansion  $R\propto t^{1/a(1-\beta)} $, which gives the singular model and the expansion is driven by particle creation. We observe that the effect of viscous coefficient becomes negligible. The model shows singularity and the expansion continuously slows down but never reverses. Thus, we see that this model evolves from a de Sitter state as  $t\to -\infty $  to zero curvature Friedmann state as  $t\to +\infty $. At $t=0$, the scale factor and energy density are finite, where as the scale factor tends to zero and the energy density becomes infinite as  $t\to -\infty $. The viscosity removes the initial singularity at finite past i.e. moves it to the infinite past. The energy density and bulk viscosity decrease as time passes. Similarly, for radiation -dominated phase where the bulk viscosity dominates we get $R\propto \exp(Ht)$, where $H=4(1-\beta)/9\zeta_{0}$. In the absence of viscous term, we have $R \propto t^{1/2(1-\beta)}$, $\rho \propto t^{-2}$ and $\zeta \propto t^{-2}$. The deceleration parameter has the value  $q=1$, which shows the deceleration of the universe. \\

\noindent \textbf{3.2.2 Solution with $ m < 1/2$}\\

In this case, Eqs.(44) and (49) for inflationary and radiation phases can respectively be rewritten as
\begin{equation}
H=\left[C_{1}(R/R_{*})^{-a(1-2m)(1-\beta)}+(3^{(m+1)}\zeta_{0}/2a(1-\beta))\right]^{\frac{1}{1-2m}}.
\end{equation}
\begin{equation}
H=\left[C_{1}(R/R_{*})^{-2(1-2m)(1-\beta)}+(3^{(m+1)}\zeta_{0}/4(1-\beta))\right]^{\frac{1}{1-2m}}.
\end{equation}
\noindent In this case, the universe evolves into a viscosity dominated steady state era. If the coefficient of bulk viscosity decays sufficiently slowly, the late epochs of the universe will be viscosity dominated, and the universe will enter a final inflationary era with steady state characteristic. This case is distinct from Murphy's case or other values of  $m$  with  $2m>1$. This can also be explained from (44) and (49). If the coefficient of bulk viscosity decays slowly, i.e. if  $3^{m+1} \zeta _{0} /2a>>C_{1}(R/R_{*} )^{a(2m-1)} $, there is exponential expansion  $R\propto \exp (H_{0} t)$. Thus, at any finite proper time in the past, the curvature is finite. The viscosity has removed the initial singularity. On the other hand if   $C_{1}(R/R_{*} )^{a(2m-1)} >>3^{m+1} \zeta _{0} /2a$, we get power --law expansion  $R\propto t^{1/a(1-\beta)} $  and the effects of viscosity are negligible. Similarly, we can discuss the behavior of solution in radiation -dominated phase.\\

\noindent \textbf{3.2.3 Solution with $ m = 1/2$}\\

The solutions obtained in subcases (3.2.1) and (3.2.2) are not valid for $m=1/2$. Let us consider the case $m=1/2$ and in this case $\zeta =\zeta_{0}\rho^{1/2}$. Now, Eq.(42) becomes
\begin{equation}
H'+\left[\frac{3(1-\beta)\gamma (R)}{2} -\alpha \right]\frac{H}{R} =0.
\end{equation}
\noindent where  $\alpha =(3\sqrt{3} \zeta _{0} /2)$ . Substituting (17) into (58) and integrating, we have the solution for Hubble parameter \\
\begin{equation}
 H=\frac{C_{2}R^{\alpha } }{[A(R/R_{*} )^{2} +(R/R_{*} )^{a}]^{(1-\beta)}}.
 \end{equation}
 \noindent where $C_{2}$ is the constant of integration. If $H=H_{*}$ for $R=R_{*}$, a relation between constants is given by
 \begin{equation}
 H_{*}=\frac{C_{2}R^{\alpha}_{*}}{(1+A)^{(1-\beta)}}.
 \end{equation}
Equation (59) further integrated to give
\begin{equation}
\int \frac{[A(R/R_{*} )^{2} +(R/R_{*} )^{a}]^{(1-\beta)}}{R^{\alpha+1}}dR=C_{2}t
\end{equation}
The constant of integration has been taken as zero for simplicity. The scale factor for inflationary and radiation -dominated phases are, respectively given by
\begin{equation}
R^{[a(1-\beta)-\alpha]}=R^{a(1-\beta)}_{*}\left[\{a(1-\beta)-\alpha\}C_{2}t\right],
\end{equation}
\noindent and
\begin{equation}
R^{[2(1-\beta)-\alpha]}=\left(\frac{R^{2}_{*}}{A}\right)^{(1-\beta)}\left[\{2(1-\beta)-\alpha\}C_{2}t\right],
\end{equation}
The corresponding Hubble parameters are given by
\begin{equation}
H=\frac{1}{[a(1-\beta)-\alpha]}\frac{1}{t}
\end{equation}
\noindent and
\begin{equation}
H=\frac{1}{[2(1-\beta)-\alpha]}\frac{1}{t}
\end{equation}
\noindent The above solutions show the power-law expansion of the universe for $a(1-\beta)> \alpha$ or $2(1-\beta)> \alpha$. The energy density, particle creation pressure and coefficient of bulk viscosity for inflationary phase are respectively given by
\begin{equation}
\rho=\frac{3}{[a(1-\beta)-\alpha]}\frac{1}{t^{2}},
\end{equation}
\begin{equation}
p_{c}=-\frac{2a\beta}{[a(1-\beta)-\alpha]}\frac{1}{t^{2}},
\end{equation}
\noindent and
\begin{equation}
\zeta=\frac{\sqrt{3}\zeta_{0}}{[a(1-\beta)-\alpha]}\frac{1}{t},
\end{equation}
\noindent Similarly, the solutions of $\rho$, $p_{c}$ and $\zeta$ for radiation dominated phase have the following forms
\begin{equation}
\rho=\frac{3}{[2(1-\beta)-\alpha]}\frac{1}{t^{2}},
\end{equation}
\begin{equation}
p_{c}=-\frac{4\beta}{[2(1-\beta)-\alpha]}\frac{1}{t^{2}},
\end{equation}
\noindent and
\begin{equation}
\zeta=\frac{\sqrt{3}\zeta_{0}}{[2(1-\beta)-\alpha]}\frac{1}{t},
\end{equation}
\noindent Singular solutions are obtained for $n=1/2$ with power-law expansion in both phases. The evolution begins from singularity with a Friedmann leading behavior and increases in size in the course of time with a decrease of energy density. The bulk viscous coefficient also decreases with time. \\
\indent From (59), a unified expression of the deceleration parameter for both inflationary and radiation phases can be expressed as a function of scale factor as
\begin{equation}
q=\frac{(1-\alpha -2\beta)A(R/R_{*} )^{2} +[a(1-\beta)-1-\alpha ](R/R_{*} )^{a} }{[A(R/R_{*} )^{2} +(R/R_{*} )^{a} ]}
 \end{equation}
\noindent Therefore, $q$ varies from  $q=[a(1-\beta)-\alpha -1]$  for  $R < < R_{*} $  to  $q=(1-\alpha -2\beta)$ for radiation phase. The deceleration parameter is positive for  $[a(1-\beta)-\alpha -1]>0$, negative for  $[a(1-\beta)-\alpha -1]<0$  and  $q=0$  for  $a(1-\beta)-\alpha =1$ during inflationary phase.  Similarly, in radiation -dominated phase the universe decelerates for $2(1-\beta)-\alpha-1 > 0$ and accelerates for $2(1-\beta)-\alpha-1 \leq 0$. \\
\indent We now study the solution in limit  $a\to 0$. In this case, Eq. (59) becomes \\
\begin{equation}
H=\frac{C_{2}R^{\alpha } }{[A(R/R_{*} )^{2} +1]^{(1-\beta)}},
\end{equation}
\noindent which can be written as
\begin{equation}
\int \frac{[A(R/R_{*})^{2} +1 ]^{a(1-\beta)}}{R^{\alpha+1}}dR=C_{2}t
\end{equation}
The constant of integration has been taken as zero for simplicity. In the limit of very small $R$, the scale factor is given by
\begin{equation}
R^{\alpha } =-\frac{1}{\alpha C_{2}} t^{-1}.
\end{equation}
we observe that  $\alpha >0$  and $C_{2}>0$  lead to contraction. As  $t\to -\infty $ , we find that  $R\to 0$. The model starts from infinite past with zero proper volume. Thus, for $a=0$  the universe is infinitely old and we have inverse power --law. Again, the radiation dominated phase is described by the same solutions as obtained for $0 < a <1$.\\
\indent From Eq.(73), a unified expression for deceleration parameter can be given in terms of scale factor as \\
\begin{equation}
q=\frac{(1-\alpha-2\beta )A(R/R_{*} )^{2} -(1+\alpha )}{A(R/R_{*} )^{2} +1}.
\end{equation}
\noindent Therefore, $q$ varies from $q=-(1+\alpha)$ for inflationary phase to $q=(1-2\beta -\alpha )$ for $R>>R_{*}$ as expected.\\

\noindent \textbf{4. Conclusion}\\

\noindent We have discussed the role of bulk viscosity in the framework of the particle creation mechanism in homogeneous and isotropic  flat FRW model. Analytical solutions for the scale factor, energy density, particle creation pressure, bulk viscous coefficient and number particle density have been derived for two early phases of the evolution of the universe with varying equation of state parameter $\gamma$. We have found that some of the models do not have the initial singularity due to the exponential inflation or even superinflation expansions. However, some models have the singular solutions.\\
\indent The concept of viscosity term has been used in a generalized form. We have analyzed the consequence of the inclusion of such a dissipative term in both inflationary and radiation --dominated phases. The solution of the models are qualitatively similar in both phases. We have observed that the bulk viscosity change the behavior from that of the perfect fluid model with particle creation. We have seen that the introduction of viscosity term in to the equation of Friedmann cosmology does not exclude automatically the appearance of singularity. Within the discussed class of models, singular and non --singular solutions appear. \\
\indent It is evident from the three cases discussed in section 3.2 that for $2m>1$ and $2m<1$, the strong energy condition $\rho+3\overline{p}\geq 0$, where $\overline{p}=p+p_{c}-3\zeta H$, is not satisfied for $C_{1}\leq 0$. It means that except for $C_{1}$ positive the energy conditions are violated throughout the evolution. But for $C_{1}>0$ and $2m<1$ the energy condition , which is equivalent to $\ddot R<0$ is satisfied at the initial stage of expansion and violated at later stages, whereas for $C_{1}>0$ and $2m>1$ the energy condition is violated at the beginning and satisfied at later stages of evolution. When $2m=1$, the solution is (59), where there is no de Sitter exact solution and the general solution is a power --law form, all energy conditions are satisfied. In this case there is a big --bang type singularity, where the evolution begins from singularity with a Friedmann leading behavior. In case subsection 3.2.2, the weak energy condition  $\rho \geq 0$  and dominant energy condition  $\rho +\overline{p}\geq 0$, are always obeyed, but the strong energy condition  $\rho +3\overline{p}\geq 0$ is violated at early times, which allows the avoidance of singularity. Therefore, there is no problem with the Hawking -Penrose $^{42}$ energy conditions for $k=0$ and $2m=1$, because it is satisfied throughout the evolution. But the energy conditions are satisfied for part of this period either at initial stages or at later stages according as $2m<1$ or $2m>1$. We also observe that similar type of energy condition $d \overline{p}/d \rho < 1$ is violated at later stages or initial stages of evolution according as $2m>1$ or $2m<1$ , but there is no problem with $2m=1$.\\
\indent In the case of constant coefficient of bulk viscosity the solutions to the field equations can be expressed in an exact exponential form. The universe starts from a non --singular state, characterized by constant and finite initial values of  $R$ , $H$  and  $\rho $. In case where the coefficient of bulk viscosity is proportional to energy density, we get a result which is similar to the solution obtained by Murphy. We observe that the bulk viscosity is capable of removing the cosmological singularity and the model approach to de Sitter universe. The solution is quite different in the case when the  $\zeta \propto \rho ^{1/2} $. In this case, the solutions correspond to power --law inflation in inflationary phase where as power --law expansion in radiation --dominated phase. The energy density and bulk viscosity decrease and tend to zero for large time. The universe starts from a singular state. We have observed that Murphy's model is the only special case of these general set of solutions. The effect of bulk viscosity is more prominent at the beginning of the universe, where the expansion scalar and energy density are quite large.  The bulk viscosity are negligible for non-relativistic and ultra-relativistic fluids but are important for the intermediate cases. As we can analyze from this work that the effect of bulk viscosity is negligible in late time. It affects only during the early universe. In this paper the bulk viscosity  have been investigated mainly to search the non-singular models as motivated by Murphy $^{27}$ and Maartens $^{26}$. The particle creation rate increases the rate of expansion in each model and exhibits the singularity. The expansion rate depends on the parameter $\beta$. Therefore, it is thought that the negative pressure caused by matter creation may play the role of a dark energy and drives the accelerating expansion of the universe.\\

\noindent \textbf{Acknowledgement}\\
 \noindent The author expresses his sincere thanks to the referee for his constructive suggestions. The author is also thankful to Prof. M. Sami, Center for Theoretical Physics, Jamia Millia Islamia university, India for helping to improve the manuscript.\\

\noindent \textbf{References}\\
\noindent  1. I. Prigogine, J. Geheniau, E. Gunzig  and P. Nardone P,  {\it Proc. Nat. \\
\indent Acad. Sci. } {\bf85} 7428 (1988).\\
\noindent 2. I. Prigogine, J. Geheniau , E. Gunzig and P. Nardone P  {\it Gen. Relativ. \\
\indent Grav.} {\bf21} 767 (1989).\\
\noindent 3. J.A.S. Lima, M.O. Calv\~{a}o  and I. Waga, {\it Cosmology, Thermodynamics \\
\indent and Matter Creation} World Scientific Singapore {\bf317} (1991).\\
\noindent 4. M.O. Calv\~{a}o, J.A.S. Lima and I. Waga, {\it Phys. Lett. A} {\bf162} 233 (1992).\\
\noindent 5. J.A.S. Lima and A.S. Germano,  {\it Phys Lett A} {\bf170} 373 (1992).\\
\noindent 6. J.A.S. Lima, A.S. Germano  and L.R.W. Abramo,  {\it Phys. Rev. D} {\bf53} \\
\indent 4287 (1996).\\
\noindent 7. W. Zimdahl, J. Triginer  and D. Pav\'{o}n,  {\it Phys. Rev.D} {\bf54} 6101 (1996).\\
\noindent 8. V.B. Johri and K. Desikan, {\it Astro. Lett. Comm.} {\bf33} 287 (1996).\\
\noindent 9. J.A.S. Lima and J.S. Alcaniz,  {\it Astron. Astrophys.} {\bf 348} 1 (1999).\\
\noindent 10. J.S. Alcaniz and J.A.S. Lima, {\it Astron. Astrophys.} {\bf349} 72 (1999).\\
\noindent 11. W. Zimdahl, et al., {\it Phys. Rev. D} {\bf64} 063501 (2001).\\
\noindent 12. Yuan Qiang, Tong-Jiezhang and Yi. Ze-Long, {\it Astrophys. Space Sci.} \\
\indent {\bf311} 407 (2007); arXiv:astro-ph/0503123.\\
\noindent 13. T. Padmanbhan and S. M. Chitre, {\it Phys. Lett. A} {\bf120} 433 (1987).\\
\noindent 14. V.B. Johri and R. Sudarshan, {\it Phys. Lett. A} {\bf132} 316 (1988).\\
\noindent 15. J.D. Barrow, {\it Phys. Lett. B} {\bf180} 335 (1986).\\
\noindent 16. J.D. Barrow, {\it Nucl. Phys. B} {\bf310} 743 (1988).\\
\noindent 17. R. Sudarshan and V.B. Johri, {\it Gen. Relativ. Grav.} {\bf26} 41 (1994).\\
\noindent 18. I. Brevik and G. Stokkan, {\it Astrophys. Space Sci.} {\bf239} 89 (1996).\\
\noindent 19. J. Triginer and D. Pav\'{o}n, {\it Gen. Relativ. Grav.} {\bf26} 513 (1994).\\
\noindent 20. K. Desikan, {\it Gen. Relativ. Grav.} {\bf29} 435 (1997).\\
\noindent 21. G.P. Singh and A. Beesham,  {\it Aust. J. Phys.} {\bf 52} 1039 (1999).\\
\noindent 22. V.B. Johri and S.K. Pandey,  {\it Int. J. Theor. Phys.} 38 1981 (1999).\\
\noindent 23. G.P. Singh, R.V. Deshpande and T. Singh,  {\it Astrophys Space Sci.} {\bf282} \\
\indent 489 (2002).\\
\noindent 24. G.P. Singh and A.Y. Kale, {\it Astrophys. Space Sci.} {\bf331} 207 (2011).\\
\noindent 25. {\O}. Gr{\o} n,  {\it Astrophys. Space Sci}. {\bf173} 191 (1990).\\
\noindent 26. R. Maartens, {\it Class. Quantum Grav.} {\bf12} 1455 (1995).\\
\noindent 27. G.L. Murphy, {\it Phys. Rev. D} {\bf 8} 4231 (1973).\\
\noindent 28. A. Beesham,  {\it Int. J. Theor. Phys.} {\bf 25} 1295 (1986).\\
\noindent 29. A.I. Arbab, {\it Chin. Phys. Lett.} {\bf25} 4497 (2008).\\
\noindent 30. L.P. Chimento, S.J. Alejanadro and D. Pav\'{o}n, {\it Phys. Rev.D} {\bf 52}\\
\indent 063509 (2002).\\
\noindent 31. A.H. Guth, {\it Phys. Rev.} {\bf23D} 347 (1981).\\
\noindent 32. M.S. Madsen and G. F. R. Ellis, {\it Mon. Not. R. Astr. Soc.} {\bf234} 67 \\
\indent (1988).\\
\noindent 33. M. Israelit and N. Rosen, {\it Astrophys. J.} {\bf342} 627 (1989).\\
\noindent 34. M. Israelit and N. Rosen, {\it Astrophys. Space Sci.} {\bf204} 317 (1993).\\
\noindent 35. J.C. Carvalho,  {\it Int. J. Theor. Phys.} {\bf35} 2019 (1996).\\
\noindent 36. C.P. Singh, S. Kumar and A. Pradhan,  {\it Class. Quantum  Grav.} {\bf24} \\
\indent 455 (2007).\\
\noindent 37. C.P. Singh,  {\it Nuovo Cim. B} {\bf122} 89 (2007).\\
\noindent 38. C.P. Singh, {\it Pramana J. Phys.} {\bf71} 33 (2008).\\
\noindent 39. C.P. Singh and A. Beesham, {\it Astrophys. Space Sci.} {\bf336} 469 (2011).\\
\noindent 40. W. Israel and J.M. Stewart, {\it Phys. Lett. A} {\it 58} 213 (1976).\\
\noindent 41. V.A. Belinskii  and I.M. Khalatnikov, {\it Sov. Phys. JETP} {\bf 42} 205 (1976).\\
\noindent 42. S. Hawking and R. Penrose, {\it Proc. R. Soc. London Ser. A} {\bf314} 529\\
\indent (1970).\\
\end{document}